\documentclass[prb,aps,amsmath,amssymb,amsfonts,floatfix,superscriptaddress,showpacs,twocolumn]{revtex4}
\usepackage[dvips]{color}
\usepackage{amsmath}
\usepackage{amssymb}
\usepackage{epsfig}
\usepackage{epsf}
\usepackage{array}
\usepackage{color}
\usepackage{graphicx}
\usepackage{epstopdf}
\usepackage{subfigure}

\usepackage{color}

\begin{document}
\bibliographystyle{apsrev}

\title{Electron-Hole Liquids  in Transition Metal Oxide Heterostructures}
\author{A. J. Millis}
\affiliation{Department of Physics, Columbia University,
538 West 120th Street, New York, NY10027, USA}
\author{Darrell G. Schlom}
\affiliation{Department of Materials Science and Engineering, 230 Bard Hall, Cornell University Ithaca, NY  14853-1501}

\begin{abstract}
Appropriately designed transition metal oxide heterostructures involving small band gap Mott insulators  are argued to support spatially separated electron and hole gasses at equilibrium. Spatial separations and carrier densities favoring the formation of excitonic states are achievable. The excitonic states may exhibit  potentially novel properties.  Energetic estimates are given, candidate material systems are discussed, and the possibility of large photvoltaic effects is mentioned.\end{abstract}

\date{\today}


\maketitle

Electron-hole bound states are a topic of longstanding importance in condensed matter physics and play a crucial role in solar energy conversion. Excitons dynamically generated by  incident photons dominate the optical response of many materials including organic compounds such as carbon nanotubes \cite{Spataru04,Wang05,Dresselhaus07} and transition metal oxides \cite{Ideguchi08}  such as $Cu_2O$. A dense gas of excitons may bose condense or form a Wigner crystal.\cite{Hanna02,Joglekar06} Excitonic condensed states are of interest because  the excitons may be spin singlet or triplet, or have more complicated properties in the presence of spin orbit coupling \cite{Hakioglu07} and in a topological insulators\cite{Seradjeh09}, and also  couple via dipolar interactions.\cite{Balatsky04}  

Considerable effort has been invested over the years in optically generated electron-hole liquids,\cite{Butov02,Snoke07,Ideguchi08} but creating and manipulating  a sufficiently high-density optically excited particle-hole gas while preventing it from recombining has proven  challenging.   An alternative route, proposed by Zhu et. al.\cite{Zhu95} is to construct a double quantum well system in which one quantum well contains holes and the other contains electrons. The spatial separation prevents recombination while if the quantum wells are close enough the electrons and holes may interact.
\begin{figure}
\includegraphics[width=0.85\columnwidth,angle=0]{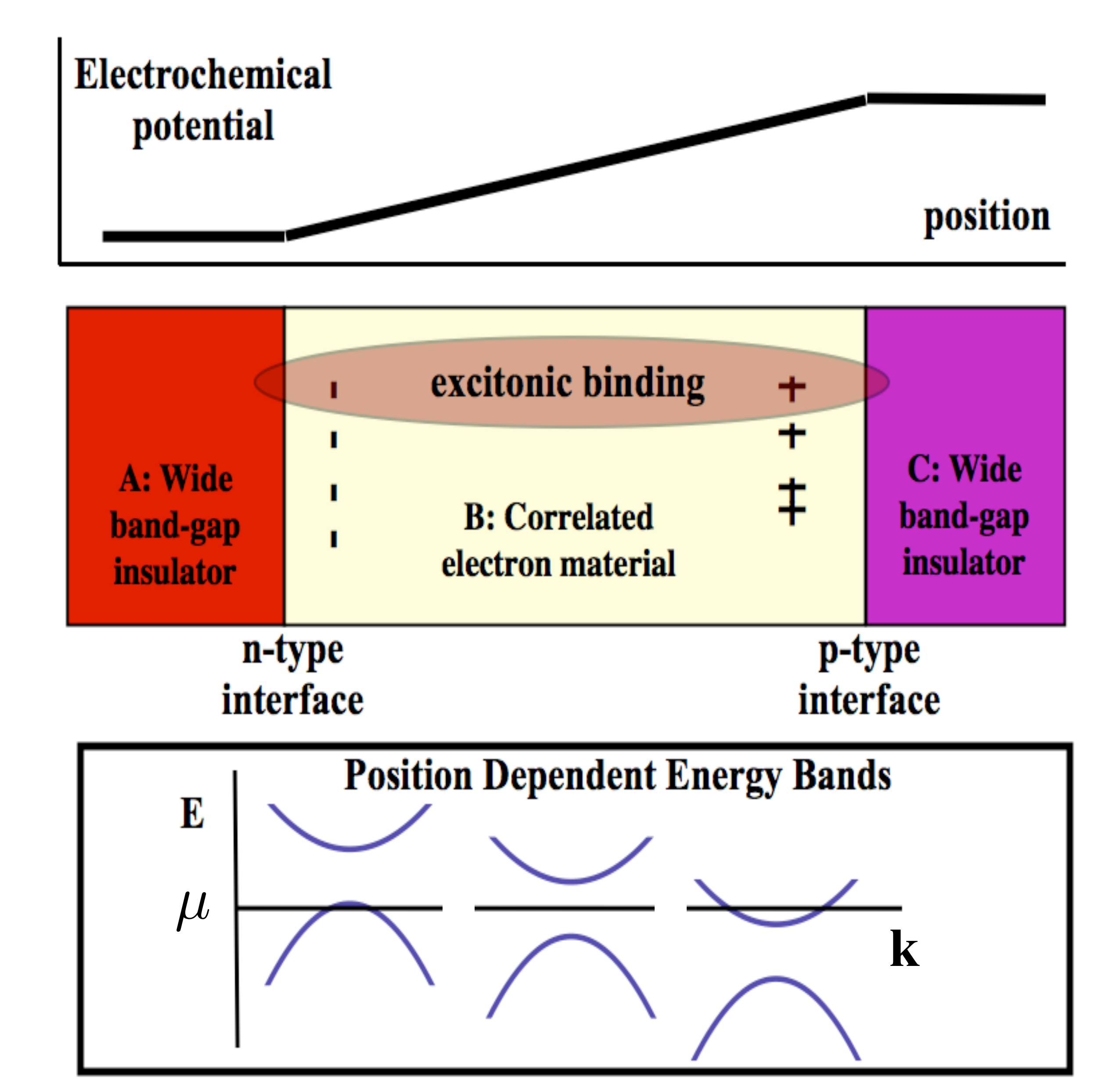}
\caption{Middle panel: sketch of heterostructure configuration in which narrow-gap insulator (for this paper, a correlated electron compound) is placed between two wide band-gap semiconductors. Left panel: potential drop across heterostructure created by polarization discontinuities at {\bf AB} and {\bf AC} interfaces. Right panel: band-bending  presented as energy $E$ vs momentum $k$ plots at three positions across the heterostructure, with chemical potential $\mu$ indicated.  }
\label{excitonfig}
\end{figure}

Double quantum wells can be fabricated in semiconductor systems such as $(Al,Ga)As$ but perhaps because of the large bandgap  it  has not been possible to bias these systems strongly enough to create significant equilibrium electron and hole populations.   Eisenstein and co-workers have created double-well systems in which each well was tuned to the $\nu=1/2$ quantized Hall state, thereby producing a bose condensate with highly unusual and still incompletely understood properties including nearly dissipationless transport characterized by counterflowing currents in the two layers.\cite{Eisenstein95,Su08} Recent theoretical papers have  raised the possibility of similar effects occurring in graphene bilayers.\cite{Min08, Kharitonov08,Lozovik09} Excitonic instabilities have also been considered in intrinsically compensated materials with coexisting electron and hole bands \cite{Brinkman72} including high-$T_c$ cuprates. \cite{Brinkman05}

In this paper we propose a different route to the formation of proximal electron and hole liquids.   The basic idea is sketched in the middle panel of Fig. [\ref{excitonfig}]: an oxide heterostructure involving a thin layer of narrow gap correlated (`Mott') insulator (labelled as $B$ in the figure) sandwiched between two possibly different wide-bandgap insulators (labelled as $A$ and $C$ in the figure). If the $AB$ and $AC$ interfaces are polar, then in the absence of charge reconstruction \cite{Hesper00,Okamoto04a} an internal electric field will be generated, leading to a potential drop  which scales linearly with the thickness of the correlated material (top panel of Fig [\ref{excitonfig}]). The potential drop causes band-bending (shown in the lower panel of Fig. [\ref{excitonfig}]) which, if large enough, pushes the conduction band  below the Fermi level on one side of the structure and the valence band above the Fermi level on the other side,  leading to electron and hole  accumulation respectively. If the band gap of the inner material (`$B$'') is smaller than the band gaps of the outer materials (``$A$'' and ``$C$'') and the heterojunctions have type I band offsets  then the electron and hole gasses will remain within layer B.  

The idea is a  variant of the proposal of Ref [\onlinecite{Zhu95}]. The importance of  polar-discontinuity fields and the possibility of exciton formation was previously noted in the context of the non-Mott $LaAlO_3/SrTiO_3$ interface by Bristowe, Artacho and Littlewood \cite{Bristowe09}. However, several features, apparently not previously noted, make the `tricolor'  oxide heterostructures  involving Mott insulators particularly attractive candidates. First, Mott  insulators often have relatively small gaps $2\Delta \sim 0.3-2eV$ \cite{Imada98} so that the electric fields required to produce the needed band-bending need not be prohibitively large. Second, the physics of correlated materials is local: the relevant length scales for charge phenomena  are of the order of a unit cell, while interlayer couplings are either intrinsically weak (as in the case of insulating parent compounds of high $T_c$ materials) or can be made to be weak by appropriately induced orbital order. Third,  the polarization discontinuities characteristic of many oxide interfaces   produce substantial  internal fields\cite{Mannhart08}, corresponding in the ideal case to sheet charge densities of one-half electron per $\sim16\AA^2$.  With a typical transition metal oxide dielectric constant of order $10$ for short-length scale phenomena  \cite{Aryasetiwan04,Aryasetiwan06,Okamoto06},  this corresponds to  an electric field of order $0.25-0.5$ $eV/\AA$, so the polar discontinuity  voltage drop across even  a few-unit-cell thick Mott channel  would create enough band-bending to establish proximal electron and hole liquids without the application of  externally applied voltages. We also note that even though most Mott and charge-transfer insulators exhibit strongly particle-hole asymetric doping properties,  charge neutrality requires that  in the absence of chemical defects the densities of electrons and holes must be  equal,  Of course, in practice the presence of defects means that  an external voltage will have to be applied to balance the system, but the voltages required need not be large. 
\begin{figure}
\includegraphics[width=0.45\columnwidth,angle=0]{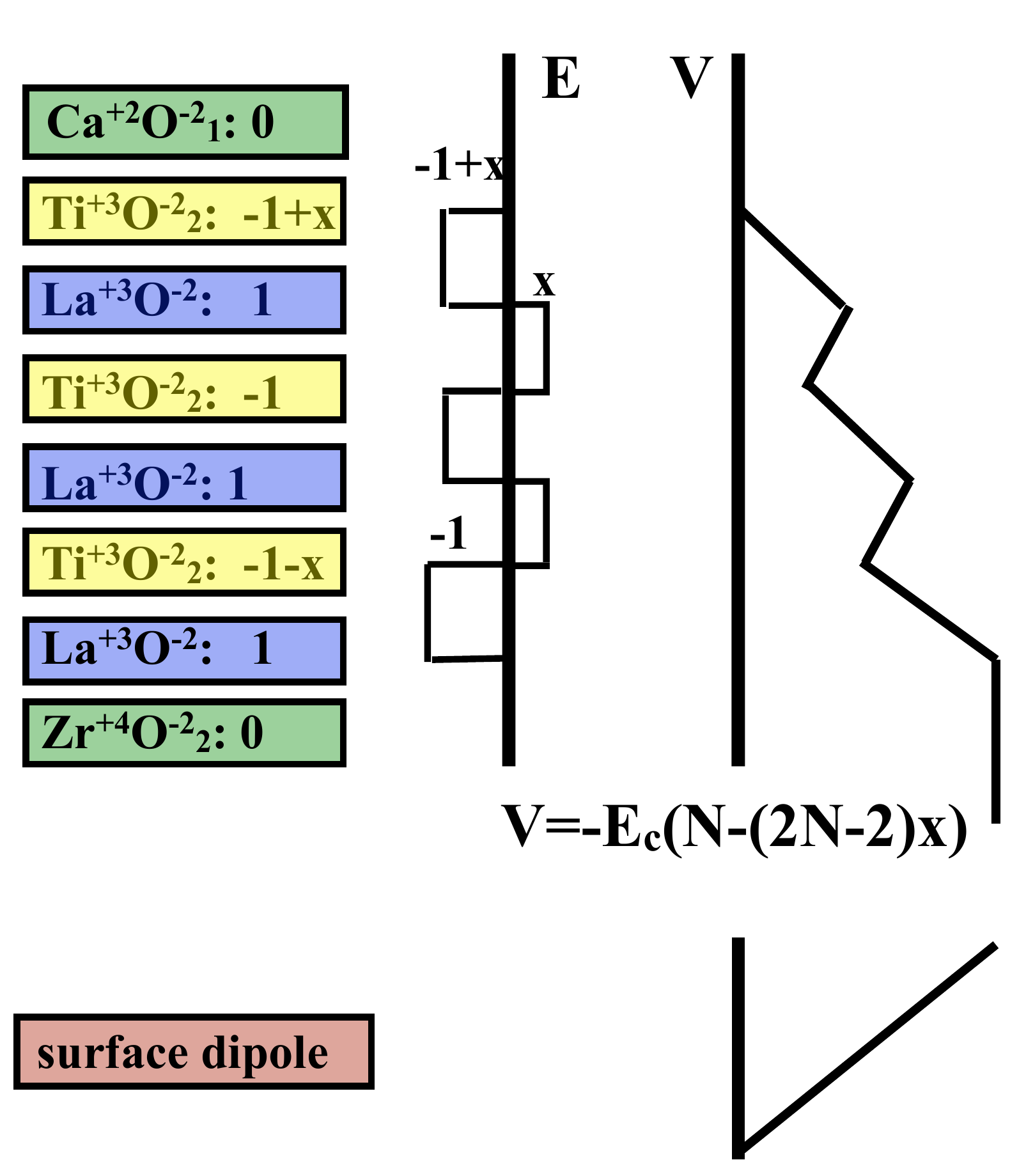}
\includegraphics[width=0.45\columnwidth,angle=0]{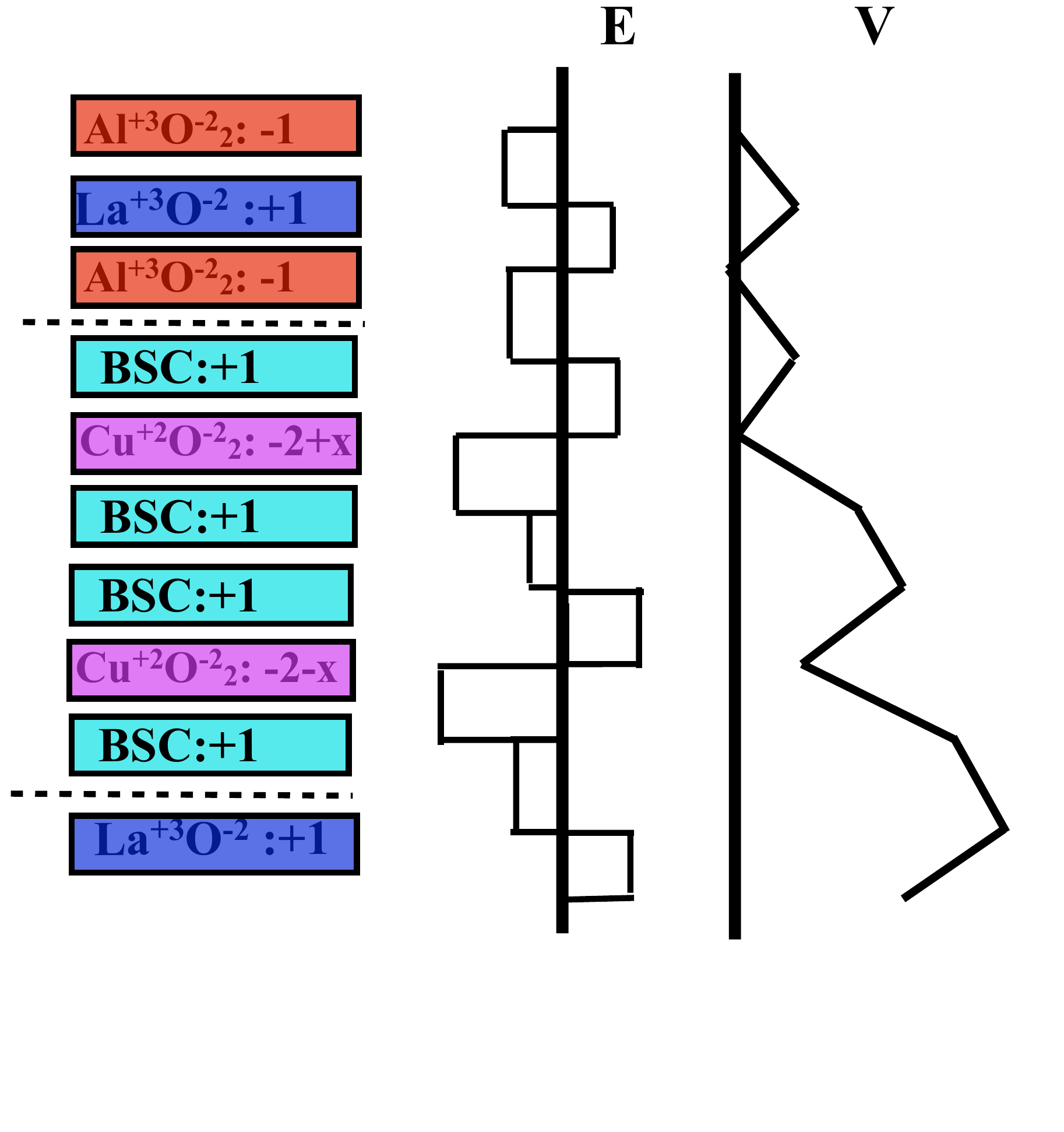}
\caption{Electrostatics of electron-hole liquid formation. Upper panel: $N=3$ layers of polar correlated insulator such as $LaTiO_3$, $LaVO_3$ or $LaCoO_3$ inserted in hypothetical  wide bandgap non-polar insulating host  (here $CaZrO_3$), with  sheet charge density (in carriers per unit cell) indicated in the ionic picture, along with the induced electric field and the electrostatic potential (in units of the energy $E_c$ defined in the main text.    Also shown is the surface dipole required for electrostatic equilibrium, Lower panel: $N=3$ layers of a non-polar correlated insulator  such as an infinite-layer cuprate inserted in a polar host, with electric field and potential drop indicated. The final surface dipole layer is not shown. }
\label{tifig}
\end{figure} 

We now use  an energy minimization argument similar to those presented by Nakagawa, Hwang and Muller\cite{Nakagawa06}  and  Bristowe, Artacho and Littlewood \cite{Bristowe09} to estimate the density of electrons and holes induced on opposite sides of an oxide heterostructure. We consider two representative   heterostructures sketched in Fig [\ref{tifig}]. The left panel shows case $P$:  a perovskite oxide with a polar $(001)$ surface embedded in a non-polar host. Specific examples might be $ReXO_3$ with $Re$ a $3^+$  rare earth such as $La$ and $X$ a transition metal such as $Ti,$, $V$, $Cr$, $Fe$, $Co$, or $Ni$. The lower panel shows case $NP$: a perovskite-related oxide with a non-polar $(001)$ surface embedded in  a host with a polar $(001)$ surface. Examples would include $La_2CuO_4$ or one of the bismuth-strontium-calcium-copper-oxide ($BSCCO$)  materials,  Viewed along the $(001)$ direction the materials consist of planes, which we idealize as having negligible thickness and a definite charge per unit area as shown in Fig [\ref{tifig}]. The potential arising from the polar discontinuity may cause the transfer of a sheet carrier density of $x$ electrons per unit cell  from one side of the structure to the other.\cite{Hesper00,Okamoto04a}. (Charge neutrality requires that the number of electrons removed from one side equals the number of electrons added to the other).  We determine $x$ by minimizing the sum of the electronic charging  energy cost to add electrons or holes to the correlated insulator   and the volume integral of the electric field energy density $E^2/(8\pi\epsilon)$.  We express  the field energy in terms of the basic scale $E_c=\frac{e^2}{\epsilon a}\frac{d}{a}$ with $a$ the in-plane lattice constant, $\epsilon$ the dielectric constant and $d$ the distance between the charge blocks shown in Fig [\ref{tifig}]; in the situations considered here $d\approx a/2$.  Using  $a \approx 4\AA$ and estimating  $\epsilon \sim10$ from previous analyses of oxide heterostructures \cite{Okamoto06} and from calculations of the screening of local interactions \cite{Aryasetiwan04,Aryasetiwan06}  we obtain $E_c\approx 0.15eV$.  For simplicity we neglect the energy cost of  the dipole layer needed to remove the potential step shown in the left panel of  Fig [\ref{tifig}]; including it would increase the density of the electron and hole gasses. We also  assume that  the transferred layer is one unit cell thick, so that $x$ electrons per  unit cell are transferred from the correlated electron layer nearest the interface on  one side of the structure to the correlated electron layer nearest the interface on the other side.    We find
\begin{eqnarray}
E_{field}^P&=&4\pi (N-1)E_c\left(1-2x+2x^2\right)
\label{Epol} \\
E_{field}^{NP}&=&4\pi (N-1)E_c\left(\frac{11}{4}-3x+3x^2\right)
\label{Enp}
\end{eqnarray}

The electronic charging energy  is the sum of  the  gap energy $\Delta_{a=el,hole}$ and the 'compressibility' energy $\kappa^{-2}=\partial \mu/\partial n $, Defining $\kappa^{-2}=(\kappa_e^{-2}+\kappa^{-2}_h)/2$ and gap  $2\Delta=\Delta_e+\Delta_h$ we get
\begin{equation}
E_a=2\Delta x+\kappa^{-2}x^2
\label{ea}
\end{equation}
$\kappa^{-2}$ is expected to be enhanced in correlated materials relative to its band theory value $\sim 0.7 eV/(unit-cell)$; an estimate\cite{Werner07doping,Dang10}, for parameters reasonable for lightly doped high-$T_c$ superconductors is $\approx 4 eV/unit-cell$. 

\begin{figure}[t]
\includegraphics[width=0.85\columnwidth,angle=0]{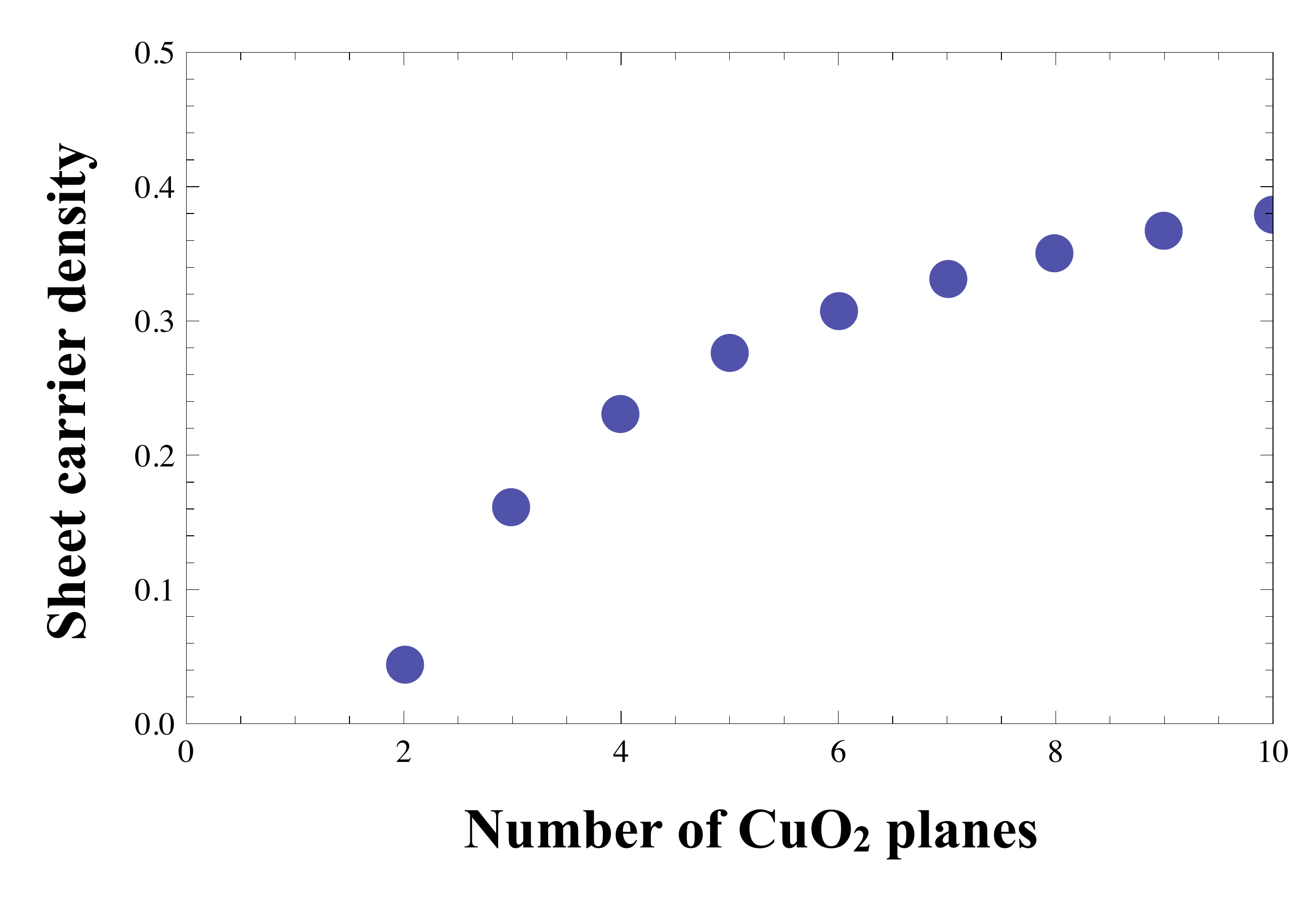}
\caption{Sheet carrier density in units of holes per in-plane unit cell as function of number of Mott insulator layers, computed using parameters appropriate to $CuO_2$ case.}
\label{densityfig}
\end{figure}

Minimizing the sum of the two terms givesgives
\begin{equation}
x=\frac{1}{2}\frac{(N-1) E_c-2\frac{\Delta}{\Lambda \pi}}{(N-1)E_c+\frac{\kappa^{-2}}{\Lambda \pi}}
\label{xofn}
\end{equation}
with  $\Lambda=4$ for the polar case and $6$ for the non-polar case. The condition $(N-1)E_c=2\Delta/(\Lambda \pi)$ is the condition that the band-bending is  larger than the gap.  The numerical estimates indicate that a  two layer system suffices to produce electron and hole gasses if the gap  $2\Delta\lesssim 2eV$.    Note also that the large value of $\kappa^{-2}$ expected in Mott insulators means that the density rises relatively slowly above its onset. An example, using the gap $2\Delta=1.75eV$ appropriate to the high-$T_c$ case, and the $E_c$ and $\kappa^{-2}$ estimates given above, is shown in Fig [\ref{densityfig}].

\begin{figure}[t]
\includegraphics[width=0.45\columnwidth,angle=0]{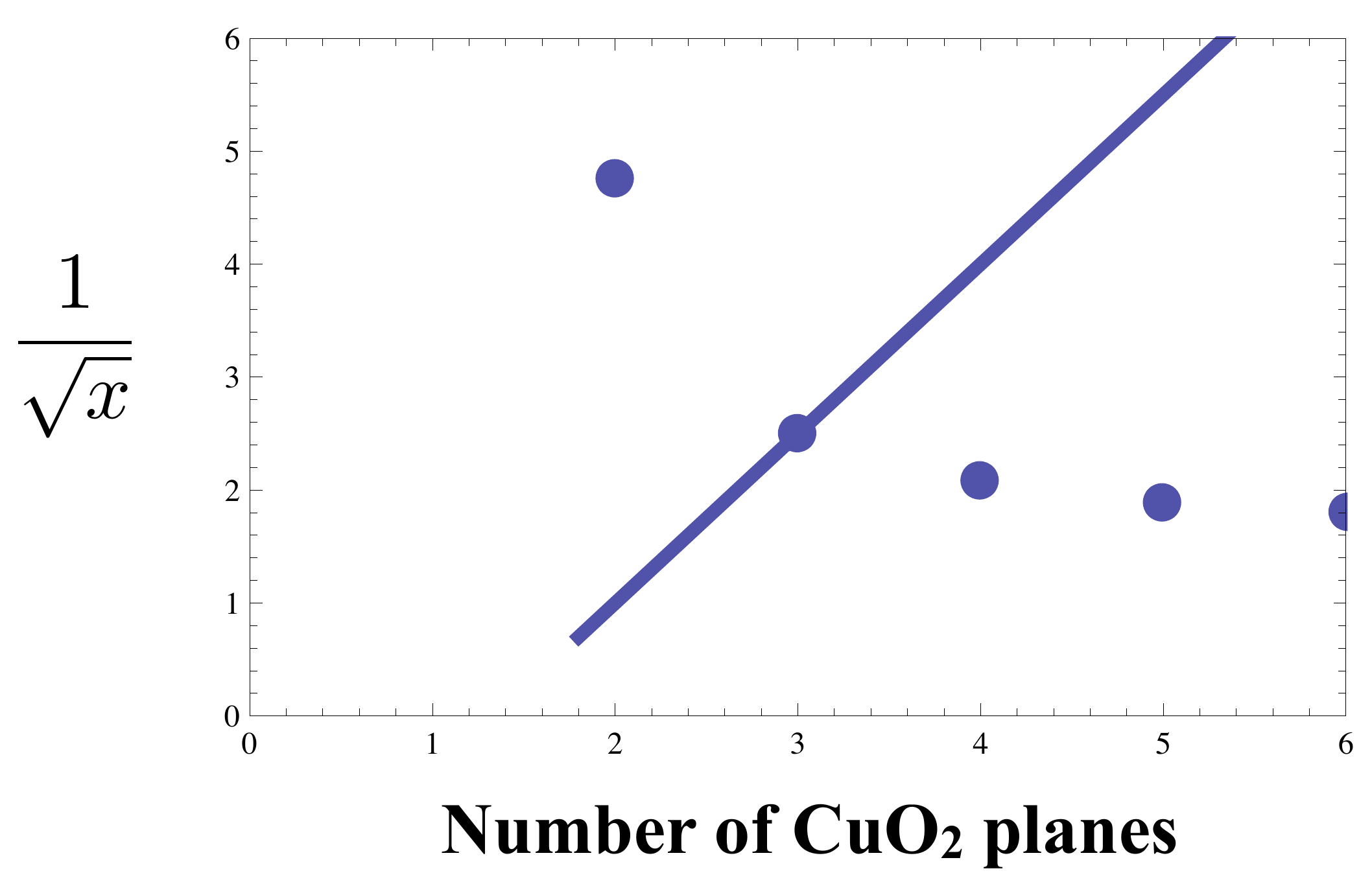}
\includegraphics[width=0.45\columnwidth,angle=0]{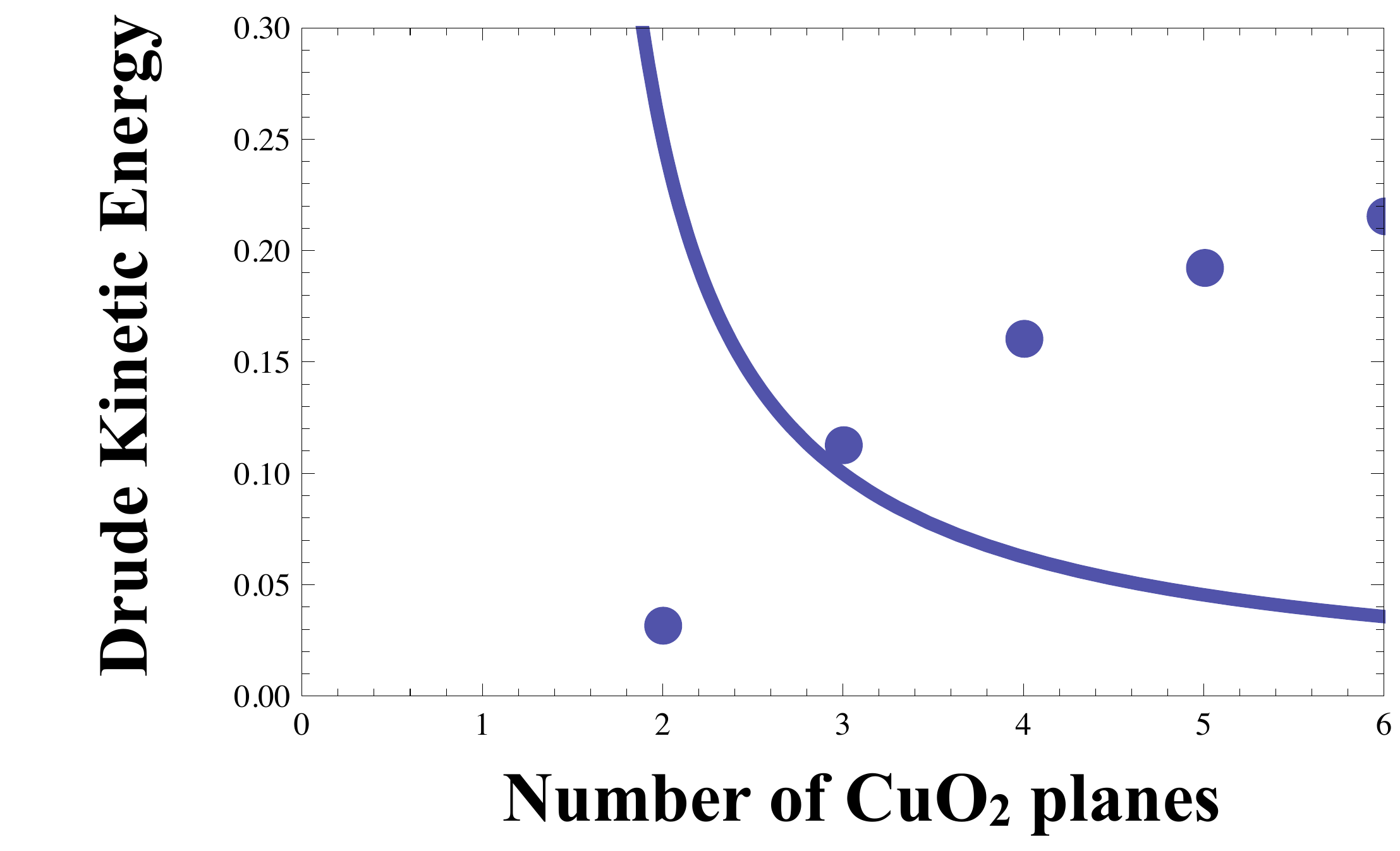}
\caption{Criteria for exciton binding. Left panel: comparison of mean interparticle spacing of electron or hole gas (solid points) computed using bandgap $2\Delta=1.75eV$ appropriate for high-$T_c$ cuprates. Solid line: interplane distance (assuming inter-block spacing is one half of in-lane lattice constant.  Right panel:  comparison of Drude kinetic energy vs doping (solid points, taken from data summarized in [\onlinecite{Comanac08}]) and Coulomb interaction at scale of interplane spacing $E_ca/d_{interplane}$.}
\label{comparisonfig}
\end{figure}

The next question is whether excitionic binding may occur.  For weakly correlated systems binding occurs if the interparticle distance in one plane is greater than to the spacing between planes, or alternatively if the binding energy is larger than  the Fermi energy.\cite{Zhu95}  The issue is more subtle in the doped Mott insulator case, because one must address the question whether one counts carriers with respect to the half filled insulator or the full and empty bands.   A detailed analysis  requires a full many-body treatment of the interplay of binding and many-body physics which is not yet available. Here we argue that the 'kinetic energy' associated with the 'Drude' (zero-frequency-centered) component of the optical conductivity gives the delocalization energy of the doped holes or electrons. For high-$T_c$ materials the Drude kinetic energy per dopant has been determined \cite{Orenstein90,Comanac08}.  Using this information and our computed electron densities we plot in the two panels of Fig [\ref{comparisonfig}] the interparticle spacing and kinetic energy, as well as the the interplane distance and the Coulombic binding energy.  One sees that $2$ or $3$ layer systems are most likely to exhibit excitonic binding.

Excitonic binding, if it occurs, may have interesting consequences. The Mott insulating channel material is likely to have a nontrivial magnetic structure, which the excitons will inherit. For example, Ribeiro et al.\cite{Ribeiro06} and Han et al \cite{Han06} modelled an electron-hole bilayer in a correlated bilayer by  coupled $t-J$ models and found  regimes of novel magnetic behavior.  But many more situations are possible.  Wigner crystallization \cite{Zhu95} and  a Wigner supersolid \cite{Joglekar06} of excitions have been theoretically predicted, and may coexist with the 'stripe' formation which is ubiquitous in lightly doped transition metal oxides.   Further, one may imagine a $d^2$ system or  a manganite-based system) where each transition metal ion is in a high-spin configuration. If the system adopts a  'layer antiferromagnetic' ground state (as occurs in some members of the 'CMR manganite family \cite{Imada98} and may with appropriate orbital order occur in other systems) and if the number $N$ of layers of correlated material is even, then in the ground state the exciton carries a spin, and exciton motion will lead to spin transport without charge transport. 

Even if excitonic binding does not occur, the properties may be of interest. The slow rise to saturation of the transferred charge  shown in Fig. [\ref{densityfig}] implies that a typical structure will have a non-negligible internal electric field. In a structure of thickness $N>2$ the intermediate layers will be in the insulating configuration. Absorption of incident light at frequencies of order the Mott gap will produce particle-hole pairs which will be rapidly dissociated by the internal field, leading (if the electron or hole layers are mobile) to a large photocurrent. If the Mott insulator is magnetic, then the electron and hole currents will each be spin polarized. 

Significant difficulties are likely to arise in attempts to put  into practice the ideas proposed here. The most serious is that while a number of reports have appeared of electron conduction at oxide interfaces \cite{Ohtomo02,Thiel06,Hotta06,Kourkotis06,Nakamura07,Mannhart08}, hole conduction seems to be very difficult to achieve  in the $ABO_3$ perovskite systems which have been the prime focus of study so far.\cite{Hotta07} Systems based on $CuO_2$ layers may be more appropriate choices. High-$T_c$ cuprate materials exist in both electron and hole-doped forms.  While most most crystal structures support either electron or hole liquids,  both $n$ type \cite{Smith91} and $p$-type \cite{Azuma92} superconductivity have been reported in  the infinite-layer material. These systems also exhibit very weak interlayer coupling which is important for the spatial segregation of the electron and hole gasses. Further, the  results in this paper are based on the assumption that the polar discontinuity fields are resolved by electronic reconstruction; of course in practice ionic and chemical effects such as interdiffusion and vacancy formation are likely to  be important. Finally, the subject  of excitonic pairing in a strongly correlated background is in its infancy. Theoretical investigation of this issue is warranted.

\acknowledgments AJM was supported by DOE-BES under grant ER-046169 and DGS acknowledges the support of ARO through grant W911NF-09-1-0415.

\bibliographystyle{apsrev}
\bibliography{excitonrefs.bib}

\begin{thebibliography}{42}
\expandafter\ifx\csname natexlab\endcsname\relax\def\natexlab#1{#1}\fi
\expandafter\ifx\csname bibnamefont\endcsname\relax
  \def\bibnamefont#1{#1}\fi
\expandafter\ifx\csname bibfnamefont\endcsname\relax
  \def\bibfnamefont#1{#1}\fi
\expandafter\ifx\csname citenamefont\endcsname\relax
  \def\citenamefont#1{#1}\fi
\expandafter\ifx\csname url\endcsname\relax
  \def\url#1{\texttt{#1}}\fi
\expandafter\ifx\csname urlprefix\endcsname\relax\def\urlprefix{URL }\fi
\providecommand{\bibinfo}[2]{#2}
\providecommand{\eprint}[2][]{\url{#2}}

\bibitem[{\citenamefont{Spataru et~al.}(2004)\citenamefont{Spataru,
  Ismail-Beigi, Benedict, and Louie}}]{Spataru04}
\bibinfo{author}{\bibfnamefont{C.}~\bibnamefont{Spataru}},
  \bibinfo{author}{\bibfnamefont{S.}~\bibnamefont{Ismail-Beigi}},
  \bibinfo{author}{\bibfnamefont{L.}~\bibnamefont{Benedict}}, \bibnamefont{and}
  \bibinfo{author}{\bibfnamefont{S.}~\bibnamefont{Louie}},
  \bibinfo{journal}{Physical Review Letters} \textbf{\bibinfo{volume}{92}},
  \bibinfo{pages}{77402} (\bibinfo{year}{2004}).

\bibitem[{\citenamefont{Wang et~al.}(2005)\citenamefont{Wang, Dukovic, Brus,
  and Heinz}}]{Wang05}
\bibinfo{author}{\bibfnamefont{F.}~\bibnamefont{Wang}},
  \bibinfo{author}{\bibfnamefont{G.}~\bibnamefont{Dukovic}},
  \bibinfo{author}{\bibfnamefont{L.}~\bibnamefont{Brus}}, \bibnamefont{and}
  \bibinfo{author}{\bibfnamefont{T.}~\bibnamefont{Heinz}},
  \bibinfo{journal}{Science} \textbf{\bibinfo{volume}{308}},
  \bibinfo{pages}{838} (\bibinfo{year}{2005}).

\bibitem[{\citenamefont{Dresselhaus et~al.}(2007)\citenamefont{Dresselhaus,
  Dresselhaus, Saito, and Jorio}}]{Dresselhaus07}
\bibinfo{author}{\bibfnamefont{M.}~\bibnamefont{Dresselhaus}},
  \bibinfo{author}{\bibfnamefont{G.}~\bibnamefont{Dresselhaus}},
  \bibinfo{author}{\bibfnamefont{R.}~\bibnamefont{Saito}}, \bibnamefont{and}
  \bibinfo{author}{\bibfnamefont{A.}~\bibnamefont{Jorio}},
  \bibinfo{journal}{Annual review of physical chemistry}
  \textbf{\bibinfo{volume}{58}}, \bibinfo{pages}{719} (\bibinfo{year}{2007}).

\bibitem[{\citenamefont{Ideguchi et~al.}(2008)\citenamefont{Ideguchi, Yoshioka,
  Mysyrowicz, and Kuwata-Gonokami}}]{Ideguchi08}
\bibinfo{author}{\bibfnamefont{T.}~\bibnamefont{Ideguchi}},
  \bibinfo{author}{\bibfnamefont{K.}~\bibnamefont{Yoshioka}},
  \bibinfo{author}{\bibfnamefont{A.}~\bibnamefont{Mysyrowicz}},
  \bibnamefont{and}
  \bibinfo{author}{\bibfnamefont{M.}~\bibnamefont{Kuwata-Gonokami}},
  \bibinfo{journal}{Phys. Rev. Lett.} \textbf{\bibinfo{volume}{100}},
  \bibinfo{pages}{233001} (\bibinfo{year}{2008}).

\bibitem[{\citenamefont{Hanna et~al.}(2002)\citenamefont{Hanna,
  D\'\i{}az-V\'elez, and MacDonald}}]{Hanna02}
\bibinfo{author}{\bibfnamefont{C.~B.} \bibnamefont{Hanna}},
  \bibinfo{author}{\bibfnamefont{J.~C.} \bibnamefont{D\'\i{}az-V\'elez}},
  \bibnamefont{and} \bibinfo{author}{\bibfnamefont{A.~H.}
  \bibnamefont{MacDonald}}, \bibinfo{journal}{Phys. Rev. B}
  \textbf{\bibinfo{volume}{65}}, \bibinfo{pages}{115323}
  (\bibinfo{year}{2002}).

\bibitem[{\citenamefont{Joglekar et~al.}(2006)\citenamefont{Joglekar, Balatsky,
  and Das~Sarma}}]{Joglekar06}
\bibinfo{author}{\bibfnamefont{Y.~N.} \bibnamefont{Joglekar}},
  \bibinfo{author}{\bibfnamefont{A.~V.} \bibnamefont{Balatsky}},
  \bibnamefont{and}
  \bibinfo{author}{\bibfnamefont{S.}~\bibnamefont{Das~Sarma}},
  \bibinfo{journal}{Phys. Rev. B} \textbf{\bibinfo{volume}{74}},
  \bibinfo{pages}{233302} (\bibinfo{year}{2006}).

\bibitem[{\citenamefont{Hakio\ifmmode~\breve{g}\else \u{g}\fi{}lu and
  \ifmmode~\mbox{\c{S}}\else \c{S}\fi{}ahin}(2007)}]{Hakioglu07}
\bibinfo{author}{\bibfnamefont{T.}~\bibnamefont{Hakio\ifmmode~\breve{g}\else
  \u{g}\fi{}lu}} \bibnamefont{and}
  \bibinfo{author}{\bibfnamefont{M.}~\bibnamefont{\ifmmode~\mbox{\c{S}}\else
  \c{S}\fi{}ahin}}, \bibinfo{journal}{Phys. Rev. Lett.}
  \textbf{\bibinfo{volume}{98}}, \bibinfo{pages}{166405}
  (\bibinfo{year}{2007}).

\bibitem[{\citenamefont{Seradjeh et~al.}(2009)\citenamefont{Seradjeh, Moore,
  and Franz}}]{Seradjeh09}
\bibinfo{author}{\bibfnamefont{B.}~\bibnamefont{Seradjeh}},
  \bibinfo{author}{\bibfnamefont{J.~E.} \bibnamefont{Moore}}, \bibnamefont{and}
  \bibinfo{author}{\bibfnamefont{M.}~\bibnamefont{Franz}},
  \bibinfo{journal}{Phys. Rev. Lett.} \textbf{\bibinfo{volume}{103}},
  \bibinfo{pages}{066402} (\bibinfo{year}{2009}).

\bibitem[{\citenamefont{Balatsky et~al.}(2004)\citenamefont{Balatsky, Joglekar,
  and Littlewood}}]{Balatsky04}
\bibinfo{author}{\bibfnamefont{A.~V.} \bibnamefont{Balatsky}},
  \bibinfo{author}{\bibfnamefont{Y.~N.} \bibnamefont{Joglekar}},
  \bibnamefont{and} \bibinfo{author}{\bibfnamefont{P.~B.}
  \bibnamefont{Littlewood}}, \bibinfo{journal}{Phys. Rev. Lett.}
  \textbf{\bibinfo{volume}{93}}, \bibinfo{pages}{266801}
  (\bibinfo{year}{2004}).

\bibitem[{\citenamefont{Lai et~al.}(2002)\citenamefont{Lai, Ivanov, Gossard,
  and Chemla}}]{Butov02}
\bibinfo{author}{\bibfnamefont{C.~W.} \bibnamefont{Lai}},
  \bibinfo{author}{\bibfnamefont{A.~L.} \bibnamefont{Ivanov}},
  \bibinfo{author}{\bibfnamefont{A.~C.} \bibnamefont{Gossard}},
  \bibnamefont{and} \bibinfo{author}{\bibfnamefont{D.~S.}
  \bibnamefont{Chemla}}, \bibinfo{journal}{Nature}
  \textbf{\bibinfo{volume}{417}}, \bibinfo{pages}{47} (\bibinfo{year}{2002}).

\bibitem[{\citenamefont{Snoke and Kavokin}(2007)}]{Snoke07}
\bibinfo{author}{\bibfnamefont{D.}~\bibnamefont{Snoke}} \bibnamefont{and}
  \bibinfo{author}{\bibfnamefont{A.}~\bibnamefont{Kavokin}},
  \bibinfo{journal}{Solid State Communications} \textbf{\bibinfo{volume}{144}},
  \bibinfo{pages}{357 } (\bibinfo{year}{2007}), ISSN \bibinfo{issn}{0038-1098}.

\bibitem[{\citenamefont{Zhu et~al.}(1995)\citenamefont{Zhu, Littlewood,
  Hybertsen, and Rice}}]{Zhu95}
\bibinfo{author}{\bibfnamefont{X.}~\bibnamefont{Zhu}},
  \bibinfo{author}{\bibfnamefont{P.~B.} \bibnamefont{Littlewood}},
  \bibinfo{author}{\bibfnamefont{M.~S.} \bibnamefont{Hybertsen}},
  \bibnamefont{and} \bibinfo{author}{\bibfnamefont{T.~M.} \bibnamefont{Rice}},
  \bibinfo{journal}{Phys. Rev. Lett.} \textbf{\bibinfo{volume}{74}},
  \bibinfo{pages}{1633} (\bibinfo{year}{1995}).

\bibitem[{\citenamefont{Eisenstein et~al.}(1995)\citenamefont{Eisenstein,
  Pfeiffer, and West}}]{Eisenstein95}
\bibinfo{author}{\bibfnamefont{J.~P.} \bibnamefont{Eisenstein}},
  \bibinfo{author}{\bibfnamefont{L.~N.} \bibnamefont{Pfeiffer}},
  \bibnamefont{and} \bibinfo{author}{\bibfnamefont{K.~W.} \bibnamefont{West}},
  \bibinfo{journal}{Phys. Rev. Lett.} \textbf{\bibinfo{volume}{74}},
  \bibinfo{pages}{1419} (\bibinfo{year}{1995}).

\bibitem[{\citenamefont{Su and MacDonald}(2008)}]{Su08}
\bibinfo{author}{\bibfnamefont{J.-J.} \bibnamefont{Su}} \bibnamefont{and}
  \bibinfo{author}{\bibfnamefont{A.~H.} \bibnamefont{MacDonald}},
  \bibinfo{journal}{Nature Physics} \textbf{\bibinfo{volume}{4}},
  \bibinfo{pages}{799} (\bibinfo{year}{2008}).

\bibitem[{\citenamefont{Min et~al.}(2008)\citenamefont{Min, Bistritzer, Su, and
  MacDonald}}]{Min08}
\bibinfo{author}{\bibfnamefont{H.}~\bibnamefont{Min}},
  \bibinfo{author}{\bibfnamefont{R.}~\bibnamefont{Bistritzer}},
  \bibinfo{author}{\bibfnamefont{J.-J.} \bibnamefont{Su}}, \bibnamefont{and}
  \bibinfo{author}{\bibfnamefont{A.~H.} \bibnamefont{MacDonald}},
  \bibinfo{journal}{Phys. Rev. B} \textbf{\bibinfo{volume}{78}},
  \bibinfo{pages}{121401} (\bibinfo{year}{2008}).

\bibitem[{\citenamefont{Kharitonov and Efetov}(2008)}]{Kharitonov08}
\bibinfo{author}{\bibfnamefont{M.~Y.} \bibnamefont{Kharitonov}}
  \bibnamefont{and} \bibinfo{author}{\bibfnamefont{K.~B.}
  \bibnamefont{Efetov}}, \bibinfo{journal}{Phys. Rev. B}
  \textbf{\bibinfo{volume}{78}}, \bibinfo{pages}{241401}
  (\bibinfo{year}{2008}).

\bibitem[{\citenamefont{Lozovik and Sokolik}(2009)}]{Lozovik09}
\bibinfo{author}{\bibfnamefont{Y.~E.} \bibnamefont{Lozovik}} \bibnamefont{and}
  \bibinfo{author}{\bibfnamefont{A.~A.} \bibnamefont{Sokolik}},
  \bibinfo{journal}{Eur. Phys. J. B} \textbf{\bibinfo{volume}{73}},
  \bibinfo{pages}{195} (\bibinfo{year}{2009}).

\bibitem[{\citenamefont{Brinkman et~al.}(1972)\citenamefont{Brinkman, Rice,
  Anderson, and Chui}}]{Brinkman72}
\bibinfo{author}{\bibfnamefont{W.}~\bibnamefont{Brinkman}},
  \bibinfo{author}{\bibfnamefont{T.}~\bibnamefont{Rice}},
  \bibinfo{author}{\bibfnamefont{P.}~\bibnamefont{Anderson}}, \bibnamefont{and}
  \bibinfo{author}{\bibfnamefont{S.}~\bibnamefont{Chui}},
  \bibinfo{journal}{Phys. Rev. Lett.} \textbf{\bibinfo{volume}{28}},
  \bibinfo{pages}{961} (\bibinfo{year}{1972}).

\bibitem[{\citenamefont{Brinkman and Hilgenkamp}(2005)}]{Brinkman05}
\bibinfo{author}{\bibfnamefont{A.}~\bibnamefont{Brinkman}} \bibnamefont{and}
  \bibinfo{author}{\bibfnamefont{H.}~\bibnamefont{Hilgenkamp}},
  \bibinfo{journal}{ArXiv:cond-mat/0503368}  (\bibinfo{year}{2005}).

\bibitem[{\citenamefont{Hesper et~al.}(2000)\citenamefont{Hesper, Tjeng,
  Heeres, and Sawatzky}}]{Hesper00}
\bibinfo{author}{\bibfnamefont{R.}~\bibnamefont{Hesper}},
  \bibinfo{author}{\bibfnamefont{L.~H.} \bibnamefont{Tjeng}},
  \bibinfo{author}{\bibfnamefont{A.}~\bibnamefont{Heeres}}, \bibnamefont{and}
  \bibinfo{author}{\bibfnamefont{G.~A.} \bibnamefont{Sawatzky}},
  \bibinfo{journal}{Phys. Rev. B} \textbf{\bibinfo{volume}{62}},
  \bibinfo{pages}{16046} (\bibinfo{year}{2000}).

\bibitem[{\citenamefont{Okamoto and Millis}(2004)}]{Okamoto04a}
\bibinfo{author}{\bibfnamefont{S.}~\bibnamefont{Okamoto}} \bibnamefont{and}
  \bibinfo{author}{\bibfnamefont{A.~J.} \bibnamefont{Millis}},
  \bibinfo{journal}{Phys. Rev. B} \textbf{\bibinfo{volume}{70}},
  \bibinfo{pages}{075101} (\bibinfo{year}{2004}).

\bibitem[{\citenamefont{Bristowe et~al.}(2009)\citenamefont{Bristowe, Artacho,
  and Littlewood}}]{Bristowe09}
\bibinfo{author}{\bibfnamefont{N.}~\bibnamefont{Bristowe}},
  \bibinfo{author}{\bibfnamefont{E.}~\bibnamefont{Artacho}}, \bibnamefont{and}
  \bibinfo{author}{\bibfnamefont{P.~B.} \bibnamefont{Littlewood}},
  \bibinfo{journal}{Phys. Rev. B} \textbf{\bibinfo{volume}{80}},
  \bibinfo{pages}{045425} (\bibinfo{year}{2009}).

\bibitem[{\citenamefont{Imada et~al.}(1998)\citenamefont{Imada, Fujimori, and
  Tokura}}]{Imada98}
\bibinfo{author}{\bibfnamefont{M.}~\bibnamefont{Imada}},
  \bibinfo{author}{\bibfnamefont{A.}~\bibnamefont{Fujimori}}, \bibnamefont{and}
  \bibinfo{author}{\bibfnamefont{Y.}~\bibnamefont{Tokura}},
  \bibinfo{journal}{Rev. Mod. Phys.} \textbf{\bibinfo{volume}{70}},
  \bibinfo{pages}{1039} (\bibinfo{year}{1998}).

\bibitem[{\citenamefont{Mannhart et~al.}(2008)\citenamefont{Mannhart, Blank,
  Hwang, Millis, and Triscone}}]{Mannhart08}
\bibinfo{author}{\bibfnamefont{J.}~\bibnamefont{Mannhart}},
  \bibinfo{author}{\bibfnamefont{D.~A.} \bibnamefont{Blank}},
  \bibinfo{author}{\bibfnamefont{H.~Y.} \bibnamefont{Hwang}},
  \bibinfo{author}{\bibfnamefont{A.~J.} \bibnamefont{Millis}},
  \bibnamefont{and} \bibinfo{author}{\bibfnamefont{J.~M.}
  \bibnamefont{Triscone}}, \bibinfo{journal}{Bulletin of the Materials Research
  Society} \textbf{\bibinfo{volume}{33}}, \bibinfo{pages}{1027}
  (\bibinfo{year}{2008}).

\bibitem[{\citenamefont{Aryasetiawan et~al.}(2004)\citenamefont{Aryasetiawan,
  Imada, Georges, Kotliar, Biermann, and Lichtenstein}}]{Aryasetiwan04}
\bibinfo{author}{\bibfnamefont{F.}~\bibnamefont{Aryasetiawan}},
  \bibinfo{author}{\bibfnamefont{M.}~\bibnamefont{Imada}},
  \bibinfo{author}{\bibfnamefont{A.}~\bibnamefont{Georges}},
  \bibinfo{author}{\bibfnamefont{G.}~\bibnamefont{Kotliar}},
  \bibinfo{author}{\bibfnamefont{S.}~\bibnamefont{Biermann}}, \bibnamefont{and}
  \bibinfo{author}{\bibfnamefont{A.}~\bibnamefont{Lichtenstein}},
  \bibinfo{journal}{Phys. Rev. B} \textbf{\bibinfo{volume}{70}},
  \bibinfo{pages}{195104} (\bibinfo{year}{2004}).

\bibitem[{\citenamefont{Aryasetiawan et~al.}(200)\citenamefont{Aryasetiawan,
  Karlsson, Jepsen, and Sch\"onberger}}]{Aryasetiwan06}
\bibinfo{author}{\bibfnamefont{F.}~\bibnamefont{Aryasetiawan}},
  \bibinfo{author}{\bibfnamefont{K.}~\bibnamefont{Karlsson}},
  \bibinfo{author}{\bibfnamefont{O.}~\bibnamefont{Jepsen}}, \bibnamefont{and}
  \bibinfo{author}{\bibfnamefont{O.}~\bibnamefont{Sch\"onberger}},
  \bibinfo{journal}{Phys. Rev. B} \textbf{\bibinfo{volume}{74}},
  \bibinfo{pages}{125106} (\bibinfo{year}{200}).

\bibitem[{\citenamefont{Okamoto et~al.}(2006)\citenamefont{Okamoto, Millis, and
  Spaldin}}]{Okamoto06}
\bibinfo{author}{\bibfnamefont{S.}~\bibnamefont{Okamoto}},
  \bibinfo{author}{\bibfnamefont{A.~J.} \bibnamefont{Millis}},
  \bibnamefont{and} \bibinfo{author}{\bibfnamefont{N.~A.}
  \bibnamefont{Spaldin}}, \bibinfo{journal}{Phys. Rev. Lett.}
  \textbf{\bibinfo{volume}{97}}, \bibinfo{pages}{056802}
  (\bibinfo{year}{2006}).

\bibitem[{\citenamefont{Nakagawa et~al.}(2006)\citenamefont{Nakagawa, Hwang,
  and Muller}}]{Nakagawa06}
\bibinfo{author}{\bibfnamefont{N.}~\bibnamefont{Nakagawa}},
  \bibinfo{author}{\bibfnamefont{H.~Y.} \bibnamefont{Hwang}}, \bibnamefont{and}
  \bibinfo{author}{\bibfnamefont{D.~A.} \bibnamefont{Muller}},
  \bibinfo{journal}{Nat Mater} \textbf{\bibinfo{volume}{5}},
  \bibinfo{pages}{204} (\bibinfo{year}{2006}).

\bibitem[{\citenamefont{Werner and Millis}(2007)}]{Werner07doping}
\bibinfo{author}{\bibfnamefont{P.}~\bibnamefont{Werner}} \bibnamefont{and}
  \bibinfo{author}{\bibfnamefont{A.~J.} \bibnamefont{Millis}},
  \bibinfo{journal}{Phys. Rev. B} \textbf{\bibinfo{volume}{75}},
  \bibinfo{pages}{085108} (\bibinfo{year}{2007}).

\bibitem[{\citenamefont{Dang et~al.}(2010)\citenamefont{Dang, Gull, and
  Millis}}]{Dang10}
\bibinfo{author}{\bibfnamefont{H.~T.} \bibnamefont{Dang}},
  \bibinfo{author}{\bibfnamefont{E.}~\bibnamefont{Gull}}, \bibnamefont{and}
  \bibinfo{author}{\bibfnamefont{A.~J.} \bibnamefont{Millis}},
  \bibinfo{journal}{unpublished}  (\bibinfo{year}{2010}).

\bibitem[{\citenamefont{Orenstein et~al.}(1990)\citenamefont{Orenstein, Thomas,
  Millis, Cooper, Rapkine, Timusk, Schneemeyer, and Waszczak}}]{Orenstein90}
\bibinfo{author}{\bibfnamefont{J.}~\bibnamefont{Orenstein}},
  \bibinfo{author}{\bibfnamefont{G.~A.} \bibnamefont{Thomas}},
  \bibinfo{author}{\bibfnamefont{A.~J.} \bibnamefont{Millis}},
  \bibinfo{author}{\bibfnamefont{S.~L.} \bibnamefont{Cooper}},
  \bibinfo{author}{\bibfnamefont{D.~H.} \bibnamefont{Rapkine}},
  \bibinfo{author}{\bibfnamefont{T.}~\bibnamefont{Timusk}},
  \bibinfo{author}{\bibfnamefont{L.~F.} \bibnamefont{Schneemeyer}},
  \bibnamefont{and} \bibinfo{author}{\bibfnamefont{J.~V.}
  \bibnamefont{Waszczak}}, \bibinfo{journal}{Phys. Rev. B}
  \textbf{\bibinfo{volume}{42}}, \bibinfo{pages}{6342} (\bibinfo{year}{1990}).

\bibitem[{\citenamefont{Comanac et~al.}(2008)\citenamefont{Comanac, de' Medici,
  Capone, and Millis}}]{Comanac08}
\bibinfo{author}{\bibfnamefont{A.}~\bibnamefont{Comanac}},
  \bibinfo{author}{\bibfnamefont{L.}~\bibnamefont{de' Medici}},
  \bibinfo{author}{\bibfnamefont{M.}~\bibnamefont{Capone}}, \bibnamefont{and}
  \bibinfo{author}{\bibfnamefont{A.~J.} \bibnamefont{Millis}},
  \bibinfo{journal}{Nat Phys} \textbf{\bibinfo{volume}{4}},
  \bibinfo{pages}{287} (\bibinfo{year}{2008}).

\bibitem[{\citenamefont{Ribeiro et~al.}(2006)\citenamefont{Ribeiro, Seidel,
  Han, and Lee}}]{Ribeiro06}
\bibinfo{author}{\bibfnamefont{T.~C.} \bibnamefont{Ribeiro}},
  \bibinfo{author}{\bibfnamefont{A.}~\bibnamefont{Seidel}},
  \bibinfo{author}{\bibfnamefont{J.~H.} \bibnamefont{Han}}, \bibnamefont{and}
  \bibinfo{author}{\bibfnamefont{D.-H.} \bibnamefont{Lee}},
  \bibinfo{journal}{EPL (Europhysics Letters)} \textbf{\bibinfo{volume}{76}},
  \bibinfo{pages}{891} (\bibinfo{year}{2006}).

\bibitem[{\citenamefont{Hoon~Han and Jia}(2006)}]{Han06}
\bibinfo{author}{\bibfnamefont{J.}~\bibnamefont{Hoon~Han}} \bibnamefont{and}
  \bibinfo{author}{\bibfnamefont{C.}~\bibnamefont{Jia}},
  \bibinfo{journal}{Phys. Rev. B} \textbf{\bibinfo{volume}{74}},
  \bibinfo{pages}{075105} (\bibinfo{year}{2006}).

\bibitem[{\citenamefont{Ohtomo et~al.}(2002)\citenamefont{Ohtomo, Muller,
  Grazul, and Hwang}}]{Ohtomo02}
\bibinfo{author}{\bibfnamefont{A.}~\bibnamefont{Ohtomo}},
  \bibinfo{author}{\bibfnamefont{D.~A.} \bibnamefont{Muller}},
  \bibinfo{author}{\bibfnamefont{J.~L.} \bibnamefont{Grazul}},
  \bibnamefont{and} \bibinfo{author}{\bibfnamefont{H.~Y.} \bibnamefont{Hwang}},
  \bibinfo{journal}{Nature} \textbf{\bibinfo{volume}{419}},
  \bibinfo{pages}{378} (\bibinfo{year}{2002}).

\bibitem[{\citenamefont{Thiel et~al.}(2006)\citenamefont{Thiel, Hammerl,
  Schmehl, Schneider, and Mannhart}}]{Thiel06}
\bibinfo{author}{\bibfnamefont{S.}~\bibnamefont{Thiel}},
  \bibinfo{author}{\bibfnamefont{G.}~\bibnamefont{Hammerl}},
  \bibinfo{author}{\bibfnamefont{A.}~\bibnamefont{Schmehl}},
  \bibinfo{author}{\bibfnamefont{C.~W.} \bibnamefont{Schneider}},
  \bibnamefont{and} \bibinfo{author}{\bibfnamefont{J.}~\bibnamefont{Mannhart}},
  \bibinfo{journal}{Science} \textbf{\bibinfo{volume}{313}},
  \bibinfo{pages}{1942} (\bibinfo{year}{2006}).

\bibitem[{\citenamefont{Hotta et~al.}(2006)\citenamefont{Hotta, Wadati,
  Fujimori, Susaki, and Hwang}}]{Hotta06}
\bibinfo{author}{\bibfnamefont{Y.}~\bibnamefont{Hotta}},
  \bibinfo{author}{\bibfnamefont{H.}~\bibnamefont{Wadati}},
  \bibinfo{author}{\bibfnamefont{A.}~\bibnamefont{Fujimori}},
  \bibinfo{author}{\bibfnamefont{T.}~\bibnamefont{Susaki}}, \bibnamefont{and}
  \bibinfo{author}{\bibfnamefont{H.~Y.} \bibnamefont{Hwang}},
  \bibinfo{journal}{Applied Physics Letters} \textbf{\bibinfo{volume}{89}},
  \bibinfo{eid}{251916} (pages~\bibinfo{numpages}{3}) (\bibinfo{year}{2006}).

\bibitem[{\citenamefont{Fitting~Kourkoutis
  et~al.}(2006)\citenamefont{Fitting~Kourkoutis, Hotta, Susaki, Hwang, and
  Muller}}]{Kourkotis06}
\bibinfo{author}{\bibfnamefont{L.}~\bibnamefont{Fitting~Kourkoutis}},
  \bibinfo{author}{\bibfnamefont{Y.}~\bibnamefont{Hotta}},
  \bibinfo{author}{\bibfnamefont{T.}~\bibnamefont{Susaki}},
  \bibinfo{author}{\bibfnamefont{H.~Y.} \bibnamefont{Hwang}}, \bibnamefont{and}
  \bibinfo{author}{\bibfnamefont{D.~A.} \bibnamefont{Muller}},
  \bibinfo{journal}{Phys. Rev. Lett.} \textbf{\bibinfo{volume}{97}},
  \bibinfo{pages}{256803} (\bibinfo{year}{2006}).

\bibitem[{\citenamefont{Nakamura et~al.}(2007)\citenamefont{Nakamura, Sawa,
  Sato, Akoh, Kawasaki, and Tokura}}]{Nakamura07}
\bibinfo{author}{\bibfnamefont{M.}~\bibnamefont{Nakamura}},
  \bibinfo{author}{\bibfnamefont{A.}~\bibnamefont{Sawa}},
  \bibinfo{author}{\bibfnamefont{H.}~\bibnamefont{Sato}},
  \bibinfo{author}{\bibfnamefont{H.}~\bibnamefont{Akoh}},
  \bibinfo{author}{\bibfnamefont{M.}~\bibnamefont{Kawasaki}}, \bibnamefont{and}
  \bibinfo{author}{\bibfnamefont{Y.}~\bibnamefont{Tokura}},
  \bibinfo{journal}{Phys. Rev. B} \textbf{\bibinfo{volume}{75}},
  \bibinfo{pages}{155103} (\bibinfo{year}{2007}).

\bibitem[{\citenamefont{Hotta et~al.}(2007)\citenamefont{Hotta, Susaki, and
  Hwang}}]{Hotta07}
\bibinfo{author}{\bibfnamefont{Y.}~\bibnamefont{Hotta}},
  \bibinfo{author}{\bibfnamefont{T.}~\bibnamefont{Susaki}}, \bibnamefont{and}
  \bibinfo{author}{\bibfnamefont{H.~Y.} \bibnamefont{Hwang}},
  \bibinfo{journal}{Physical Review Letters} \textbf{\bibinfo{volume}{99}},
  \bibinfo{pages}{236805} (\bibinfo{year}{2007}).

\bibitem[{\citenamefont{Smith et~al.}(1991)\citenamefont{Smith, Manthiram,
  Zhou, Goodenought, and T.}}]{Smith91}
\bibinfo{author}{\bibfnamefont{M.~G.} \bibnamefont{Smith}},
  \bibinfo{author}{\bibfnamefont{A.}~\bibnamefont{Manthiram}},
  \bibinfo{author}{\bibfnamefont{J.}~\bibnamefont{Zhou}},
  \bibinfo{author}{\bibfnamefont{J.~B.} \bibnamefont{Goodenought}},
  \bibnamefont{and} \bibinfo{author}{\bibfnamefont{M.~J.} \bibnamefont{T.}},
  \bibinfo{journal}{Nature} \textbf{\bibinfo{volume}{351}},
  \bibinfo{pages}{541} (\bibinfo{year}{1991}).

\bibitem[{\citenamefont{Azuma et~al.}(1992)\citenamefont{Azuma, Hiroi, Takano,
  Bando, and Takeda}}]{Azuma92}
\bibinfo{author}{\bibfnamefont{M.}~\bibnamefont{Azuma}},
  \bibinfo{author}{\bibfnamefont{Z.}~\bibnamefont{Hiroi}},
  \bibinfo{author}{\bibfnamefont{M.}~\bibnamefont{Takano}},
  \bibinfo{author}{\bibfnamefont{Y.}~\bibnamefont{Bando}}, \bibnamefont{and}
  \bibinfo{author}{\bibfnamefont{Y.}~\bibnamefont{Takeda}},
  \bibinfo{journal}{Nature} \textbf{\bibinfo{volume}{356}},
  \bibinfo{pages}{775} (\bibinfo{year}{1992}).

\end{thebibliography}

\end{document}